\documentclass{emulateapj}

\newcommand{\wlam}{$W_{\lambda}$}
\newcommand{\mo}{$^{-1}$}
\newcommand{\alphacx}{$\alpha^{4.86}_{8.46}$}
\newcommand{\alphaxu}{$\alpha^{8.46}_{14.94}$}

\slugcomment{To appear in AJ.}

\shorttitle{High Resolution Maps of 1Jy BL Lacs}
\shortauthors{Rector \& Stocke}

\begin{document}


\title{High-Resolution Radio Imaging of Gravitational Lensing Candidates in the 1 Jansky BL Lac Sample}


\author{Travis A. Rector}
\affil{National Radio Astronomy Observatory, P.O. Box O, Socorro, NM 87801}
\email{trector@nrao.edu}

\and

\author{John T. Stocke}
\affil{Center for Astrophysics and Space Astronomy, University of Colorado, Boulder, CO 80309-0389}




\begin{abstract}

While BL Lacertae objects are widely believed to be highly beamed, low-luminosity
radio galaxies, many radio-selected BL Lacs have extended radio power levels and optical emission lines that are
too luminous to be low-luminosity radio galaxies.  Also, Stocke \& Rector discovered an excess of MgII absorption
systems along BL Lac sightlines compared to quasars, suggesting that gravitational lensing may be another
means of creating the BL Lac phenomenon in some cases.
We present a search for gravitationally-lensed BL Lacs with
deep, high-resolution, two-frequency VLA radio maps of seven lensing candidates from the 1 Jansky BL Lac sample. 
We find that none of these objects are resolved into an Einstein ring like B
0218+357, nor do any show multiple images of the core.  All of the lensing candidates that were resolved show a
flat-spectrum core and very unusual, steep-spectrum extended morphology that is incompatible with a multiply
lensed system.  Thus, while these observations do not rule out microlensing, no macrolensing is observed.

\end{abstract}




\section{Introduction}

BL Lacertae objects are an extreme type of Active Galactic Nuclei (AGN), whose hallmark is their
``featureless" optical spectra (rest frame \wlam\ $\leq 5$\AA\ for any emission lines present).  They are a member
of the blazar class of AGN.  And like other blazars, their observed properties are thought to due to
bulk relativisitic outflows.  It is believed that most BL Lacs are low-luminosity, \citet{fan74} class 1
radio galaxies (FR--1s) whose jet axes are oriented roughly towards the observer \citep{bla78}. 
Several pieces of evidence support this hypothesis; e.g., most BL Lacs have giant elliptical host galaxies in
rich galaxy regions \citep{wur97} consistent with FR--1 hosts, and most have a highly core-dominated radio
structure that is consistent with a beamed source
\citep{ant85,per94,urr95}.  However, there is substantial evidence that not all BL Lacs follow this simple picture,
especially within the radio-selected ``1 Jansky" (1Jy) sample of BL Lacs \citep{sti91,rec01}.  In fact,
the extended radio power levels for fully one-half of the BL Lacs in the 1Jy sample are too luminous to be FR--1s. 
These objects also have broad, moderately luminous optical emission lines (log $L = 41-43$ erg s$^{-1}$), which are
typically seen in FR--2s but not in FR--1s.  Why FR--2-like sources appear in abundance in the 1Jy BL Lac sample
but not in X-ray selected samples \citep{rec00} is as yet unknown, although \citet{rec01} discuss several possible
factors.  

One of these factors might be gravitational lensing.  \citet{sto97} find an excess of MgII systems along the
sightlines of 1Jy BL Lacs.  Ten absorption systems were found in the spectra of 37 BL Lacs; and five of them
are very strong (\wlam\ $\geq 1$\AA), which is four to five times greater than the number expected based upon
quasar sightlines, i.e., 2.5 to 3$\sigma$ greater than the expectation value from \citet{ste92}.  This excess
suggests a correlation between the presence of absorbing gas along the BL Lac sightline and the relatively
featureless spectra of BL Lac objects, as compared to quasars.  Such a correlation can be caused by
gravitational microlensing of an AGN by stars associated with the absorbing gas, presumably within an
intervening foreground galaxy. 
\citet{ost90} postulated that BL Lacs may simply be microlensed quasars.  In their model, a background quasar
is microlensed by stars within a foreground giant elliptical galaxy.  The microlensing preferentially amplifies the
compact nonthermal continuum over the extended emission lines regions, reducing the observed
\wlam\ of the emission lines.  If its continuum is sufficiently boosted by microlensing, a quasar may
appear to be a BL Lac object.  While it is now clear that the
\citet{ost90} model does not explain all BL Lac objects, it may explain some.  And indeed, evidence of
lensing exists for several BL Lacs, e.g., AO 0235+164
\citep{sti88,web00}, MS 0205.7+3509 \citep{sto95}, PKS~0537--441 \citep{rom95}, PKS~1413+135 \citep{per96}, RGB
1745+398 \citep{nil99} and three other BL Lacs in an {\it HST} snapshot survey of BL Lacs \citep{sca99}.  There is
also evidence against the lensing hypothesis for some of these objects, e.g., PKS~0537--441 \citep{fal92} and MS
0205.7+3509 \citep{fal97}.  Further, \citet{nar90} discuss several of these sources in a microlensing context
where the foreground galaxy has a gravitational potential too shallow to produce multiple images via
macrolensing.  Thus it is not known how many BL Lacs are lensed, nor how many AGN are classified as  BL Lacs as a
result of lensing.  

Unfortunately, it is difficult to test for microlensing on a case-by-case basis.  An intervening MgII
absorption system is suggestive, but not conclusive, of lensing.  Similarly, optical imaging can, and
occasionally does, discover an intervening foreground galaxy which is offset from the BL Lac
nucleus, e.g.,  \citet{hei98}; although this too is not conclusive evidence.  Further, the characteristic light
curve of a single microlensing event is likely to be blended into multiple events, thereby producing a
variability signature which is not easily differentiated from an intrinsically variable source, e.g.
\citet{tak98}; although achromatic variability is a lensing characteristic which may be used to differentiate
\citep{web00}.  While it is not conclusive evidence of microlensing, the detection of macrolensing (i.e., multiple
images) remains the only means to confirm that an object is indeed gravitationally lensed.  So, despite the
suggestive evidence alone, currently there is only one confirmed case of a lensed BL Lac: the ``smallest Einstein
ring" source B 0218+357.  The ring is very compact; it is unresolved at arcsecond-scale resolutions, but is clearly
resolved with the VLA A-array at 8.4 GHz and higher frequencies
\citep{ode92}.  This compact radio structure is typical of gravitational lenses; e.g., all the lenses currently
known in the Jodrell Bank-VLA astronomical survey have angular separations on the order of a few arcseconds or
less \citep{kin97}.

Here we present a sensitive, high-resolution imaging study of all of the gravitational lens candidates in the
1Jy BL Lac sample which can be reached by the VLA ($\delta > -20$\arcdeg).  Nearly all of the BL Lacs in the 1Jy
sample have now been mapped at high sensitivity with the VLA A-array at 1.4 GHz \citep{mur93,per94,cas99,rec01}. 
Based upon these maps lensing candidates were chosen with the following criteria.  First, sources were chosen that
have an unusual radio morphology which is suggestive of lensing, i.e., very distorted and very luminous extended
radio structure.  The BL Lacs S4~0814+425 \citep{mur93} and PKS~2131--021 \citep{rec01} are examples of objects
with compact ($\leq 10$\arcsec) radio structure that, if interpreted as classical ``triple" sources, appear to
have edge-brightened, FR--2 like lobes but with a compact, wide-angle tail morphology that is very unusual for
FR--2s.  Alternatively, the morphology of both sources could be explained as Einstein rings that are marginally
resolved.  Secondly, sources were chosen that are unresolved or marginally resolved (i.e., highly core
dominated).  Thirdly, we chose objects whose optical spectrum contains a strong MgII absorption system.  This
latter criterion selects only high-redshift objects because only MgII absorption systems with redshifts
$z \ga 0.4$ can be detected at optical wavelengths.  Naturally, B 0218+357 was excluded from the
sample since it has already been established to be a lensed system.  Seven other objects from the 1Jy BL Lac
sample meet one or more of these criteria and were observed in this study (see Table~\ref{tbl-0}).  The columns 
are: [1] the object name; [2--3] the emission and absorption line redshifts; [4] the 20cm extended radio power (in
W Hz\mo); and [5] the 20cm radio core dominance, $f \equiv P_{core} / P_{ext}$.  Values were obtained from
\citet{rec01} and references within.

\section{Observation and Reduction}

We chose to observe with the A-array with a 50 MHz bandwidth to maximize resolution while maintaining good
sensitivity to extended, steep-spectrum structure.  Each object was observed at two frequencies so that
spectral-index maps could be generated.  Spectral index maps will show whether an object is a typical core-jet
source, in which case the steep-spectrum jets and lobes will be easily differentiated from the flat-spectrum
core, or if it is a lensed object with multiple images with the same spectral index.  
Although not attempted here, polarization-sensitive observations can provide an additional test, as the
polarization vectors of a lensed image should be unaffected by the gravitational field of the lens \citep{dye92}. 
Thus, the lensed images of each source component should also have the same polarization percentage and
position angle, once corrected for Faraday rotation due to the different light paths.  One caveat for both
 spectral-index and polarization measurements is that BL Lacs are known to vary both quantities on timescales
as short as days.  Thus, in principle multiple images of an object could display different spectral
indices and polarizations due to the difference in light travel times if the variability is intrinsic and not due
to the lens.

Lensing candidates
that are resolved at 1.4 GHz and have a largest angular size greater than 5\arcsec\ were observed at 4.86
GHz (C-band) and 8.46 GHz (X-band) to resolve individual components while maintaining excellent
sensitivity to faint extended structure.  Smaller sources were observed at 8.46 GHz and 14.94 GHz (U-band)
to improve the likelihood of resolving discrete components.   The resolution of these maps are ten times
greater than maps previously available for these sources.

The seven lensing candidates from the 1Jy BL Lac sample were observed with the VLA A-array on 20 September
1999 and 24 September 1999. Four or five scans of 4--8 minute durations were made for each source at each
frequency.  Scans were spaced to optimize coverage in the $(u,v)$ plane.  The resulting maps have sensitivities
of roughly 0.1 to 1 mJy beam\mo.

Epoch 1995.2 VLA values were used to flux calibrate the maps using multiple observations of 3C 286. 
Since these sources are highly core-dominated, a point source model was assumed to start the
self-calibration process.  Phase-only self-calibration in
decreasing solution time intervals was used for the first four iterations.  Amplitude and phase
self-calibration were then used until the maximum dynamic range was achieved, usually requiring only
one or two more iterations.  The AIPS task IMAGR was used to generate the maps and clean components. 
Robust weighting (ROBUST = 0.5) was used to achieve a smaller beam FWHM with only a 10--12\% increase
in noise over natural weighting; see \citet{bri95} for an explanation.
The core flux densities were measured by fitting the core with a single Gaussian with the synthesized
beam's parameters.  The extended flux was determined by measuring the total flux density with a box
enclosing the entire source and then subtracting the core flux density.  
Throughout this paper we assume $S \propto \nu^{+\alpha}$.

In addition to the radio observations described above, {\it Hubble Space Telescope} and ground-based optical data
in the literature \citep{sca00,pur02} were searched for optical counterparts to bright spots in the radio
structure.  However, no counterparts to the structure observed in the resolved sources were detected.  If these
are indeed lensed sources, optical counterparts should be present and be sufficiently bright to distort the
observed the optical radial profiles of these sources.  However, in all cases the objects are unresolved at
optical wavelengths.

\subsection{Summary of Radio Properties}

\begin{figure}
\plotone{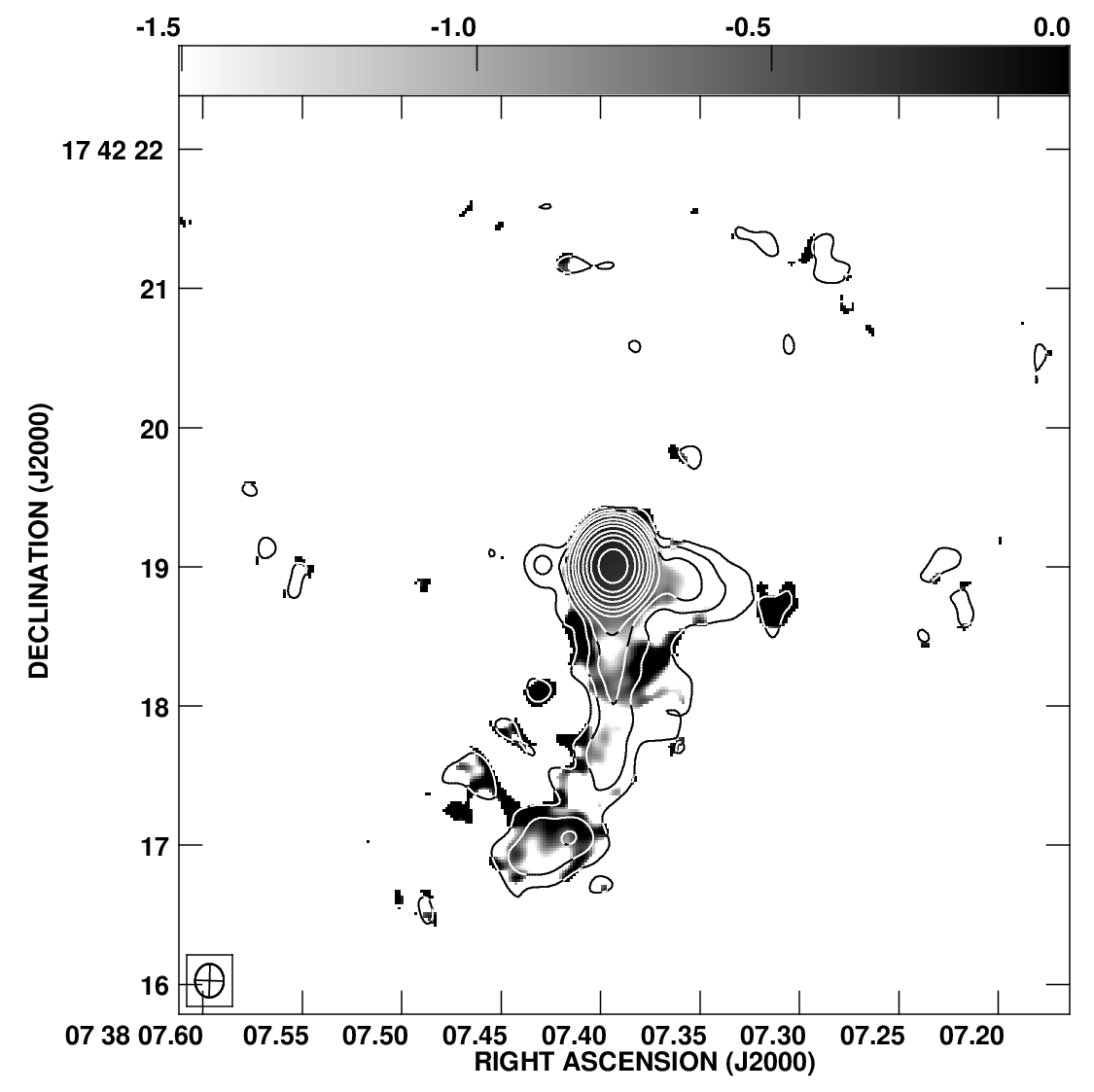}
\caption {VLA 8.46 GHz map (contour) and spectral index map (grayscale) of PKS~0735+178.  The beam is shown in
the lower left corner.  The contour levels are 0.02, 0.05, 0.1, 0.2, 0.5, 1, 2, 5, 10, 20, 50 and 100\% the peak
flux of 9.755 x 10$^{-1}$ Jy beam\mo.  The spectral index range is from $-1.5$ to $0$.}
\label{fig-1}
\end{figure}

\begin{figure}
\plotone{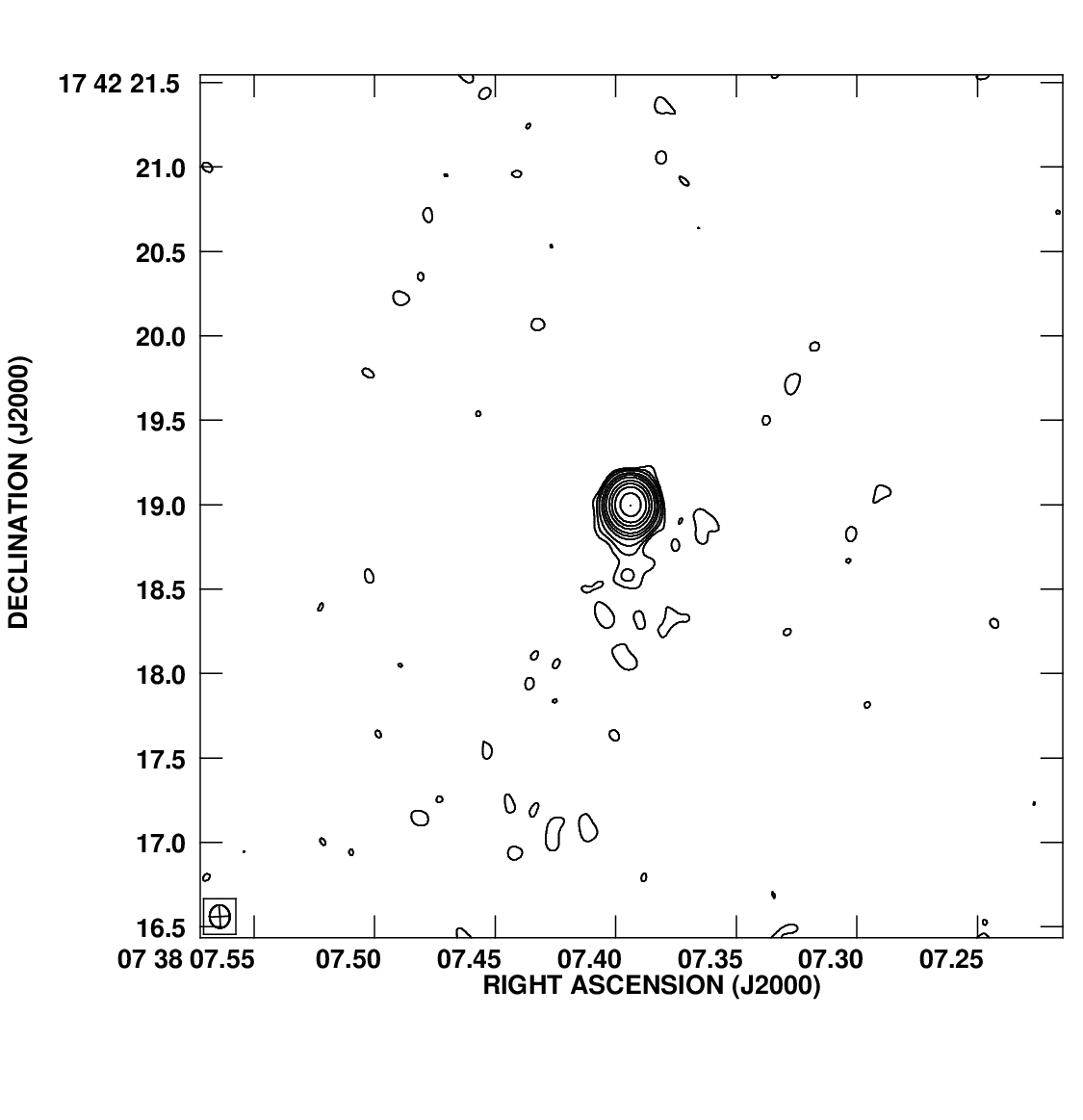}
\caption {VLA 14.94 GHz map of PKS~0735+178.  The beam is shown in the lower left corner.  The contour levels
are 0.05, 0.1, 0.2, 0.5, 1, 2, 5, 10, 20, 50 and 100\% the peak flux of 8.389 x 10$^{-1}$ Jy
beam\mo.}
\label{fig-2}
\end{figure}

\begin{figure}
\plotone{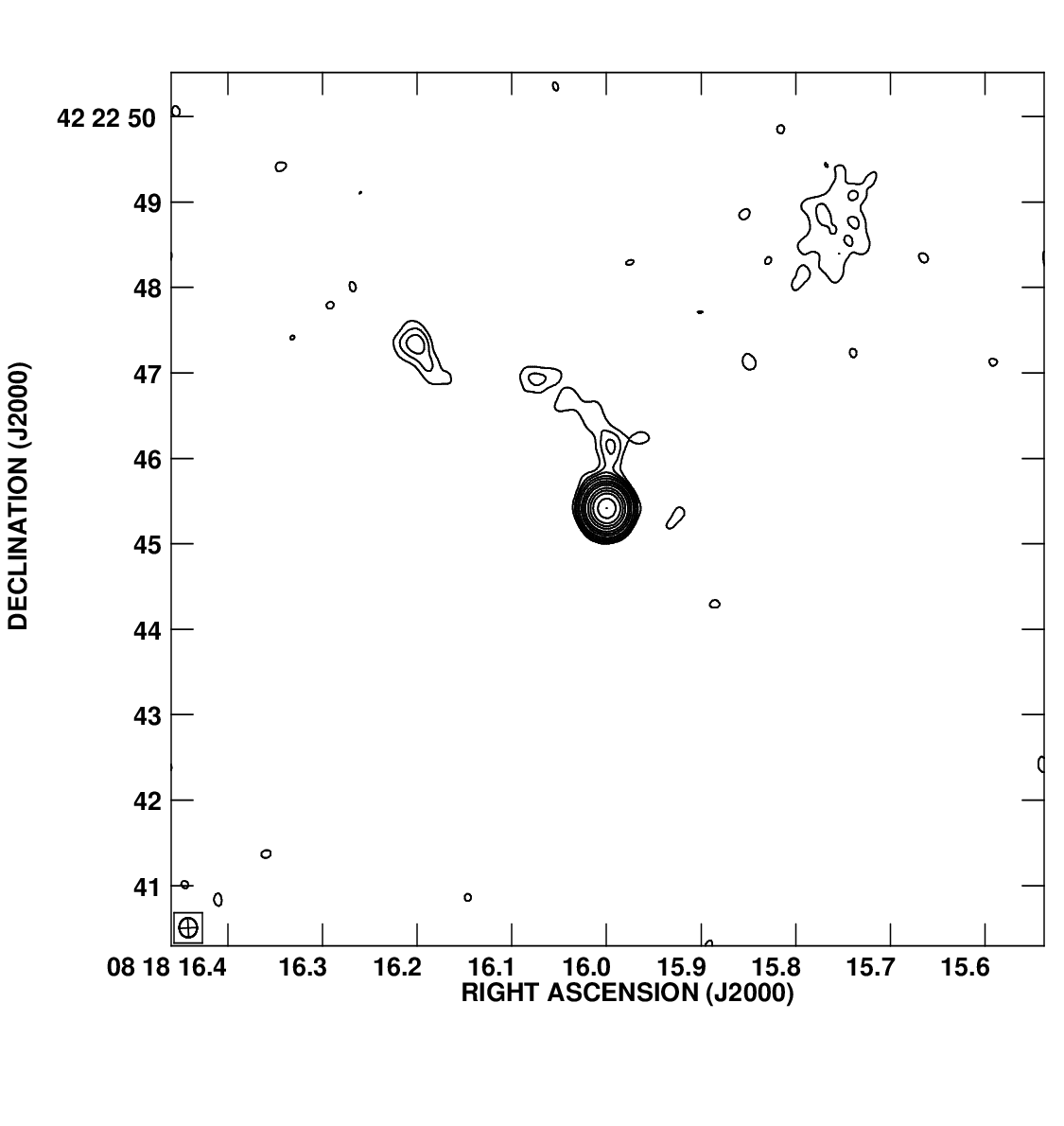}
\caption {VLA 8.46 GHz map of S4~0814+425.  The beam is shown in the lower left corner.  The contour levels
are 0.02, 0.05, 0.1, 0.2, 0.5, 1, 2, 5, 10, 20, 50 and 100\% the peak flux of 1.0723 Jy beam\mo.}
\label{fig-4}
\end{figure}

\begin{figure}
\plotone{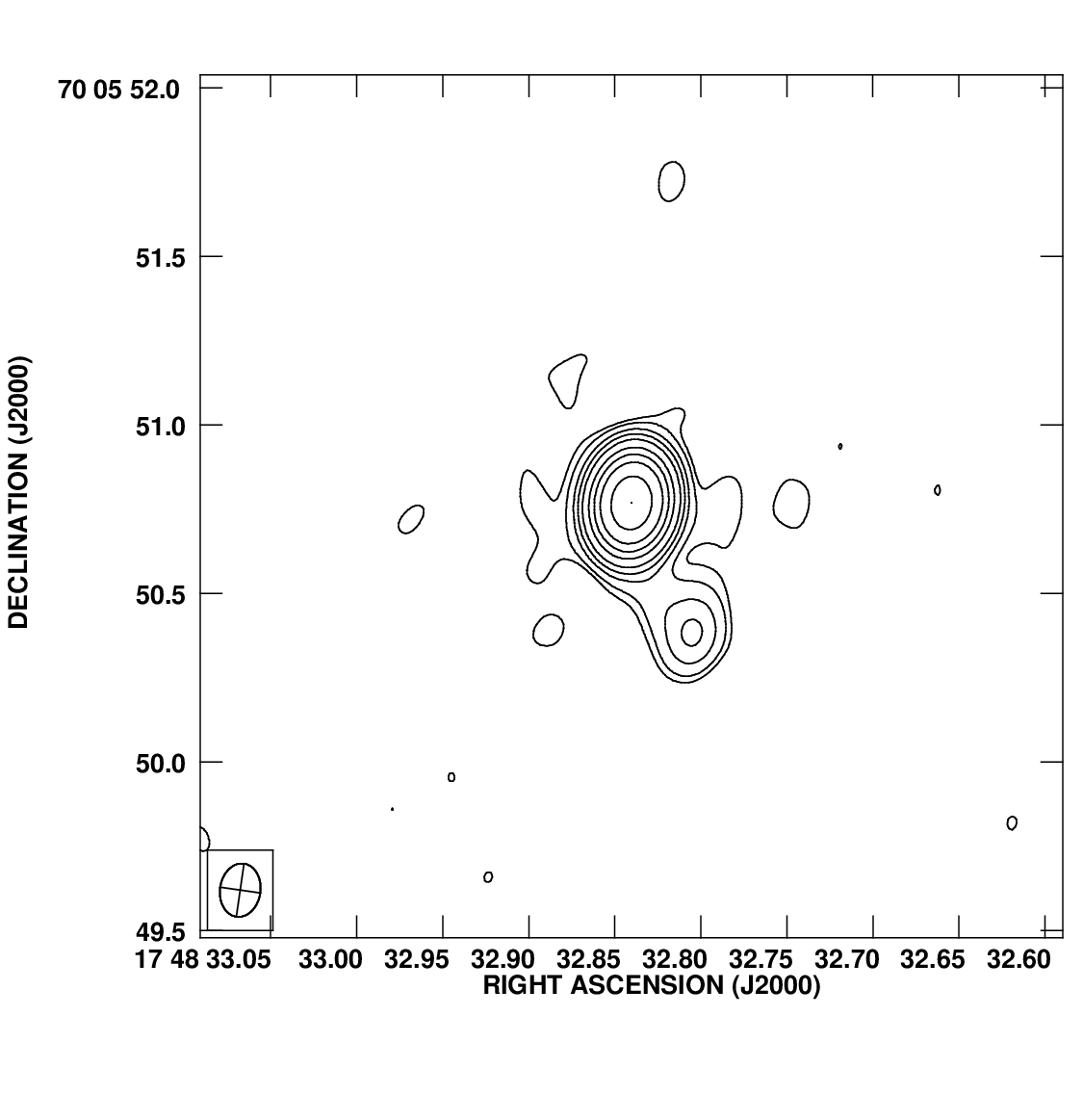}
\caption {VLA 14.94 GHz map of S4~1749+701.  The beam is shown in the lower left corner.  The contour levels
are 0.02, 0.05, 0.1, 0.2, 0.5, 1, 2, 5, 10, 20, 50 and 100\% the peak flux of 4.599 x 10$^{-1}$ Jy beam\mo.}
\label{fig-6}
\end{figure}

\begin{figure}
\plotone{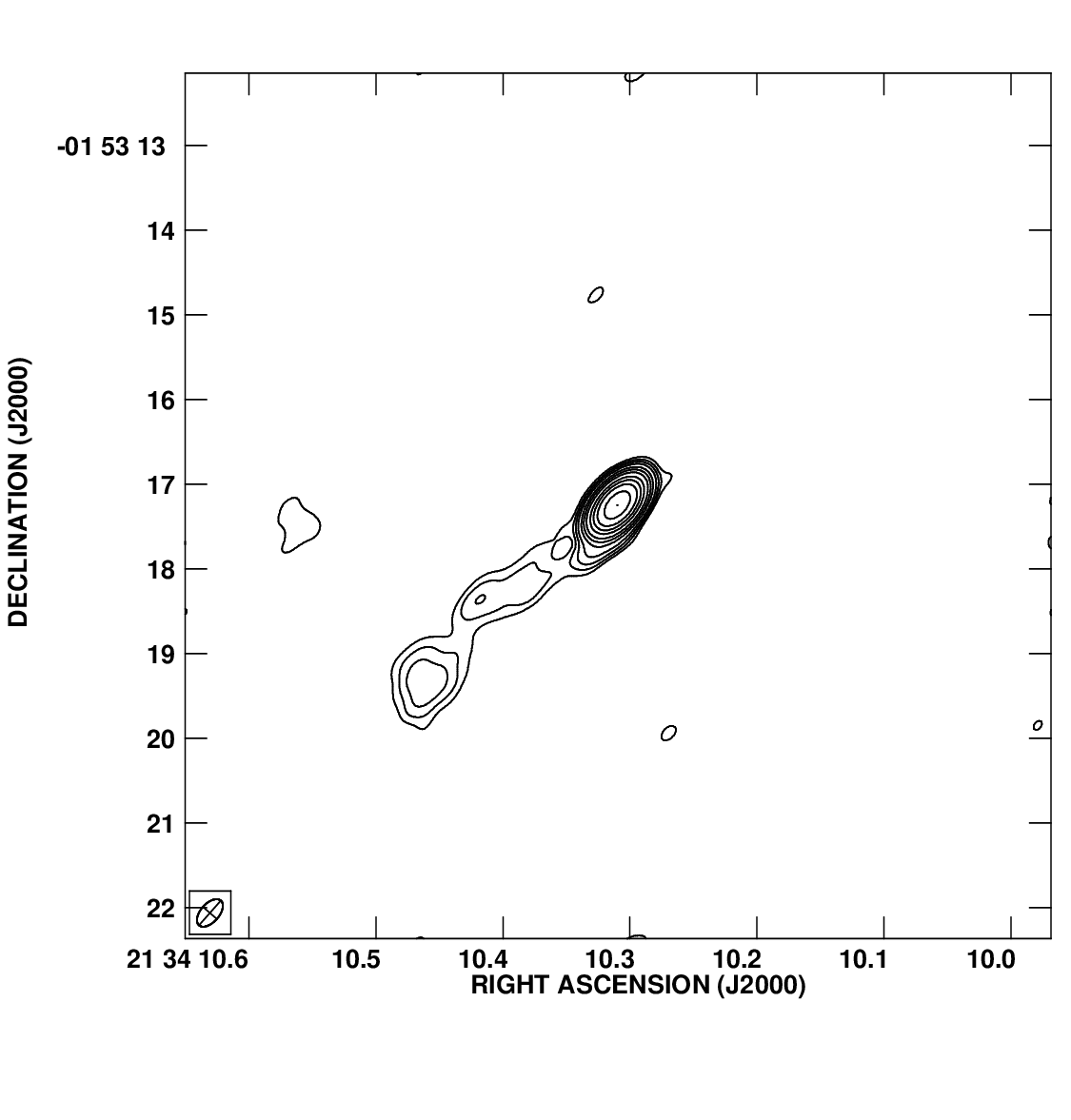}
\caption {VLA 8.46 GHz map of PKS~2131--021.  The beam is shown in the lower left corner.  The contour levels
are 0.02, 0.05, 0.1, 0.2, 0.5, 1, 2, 5, 10, 20, 50 and 100\% the peak flux of 1.8020 Jy beam\mo.}
\label{fig-8}
\end{figure}

The VLA continuum and spectral index maps are shown in Figures~\ref{fig-1} through~\ref{fig-8}.  A summary of
the radio properties is given in Table~\ref{tbl-1}.  The columns are: [1] the object name; 
[2--3] the 4.86 GHz VLA core and extended flux densities (mJy); 
[4--5] the 8.46 GHz VLA core and extended flux densities (mJy); and
[6--7] the 14.94 GHz VLA core and extended flux densities (mJy).  The error estimates for extended flux
densities are based upon the solid angular extent of the source.  For unresolved sources, an upper limit
on extended flux is conservatively estimated to be 10$\sigma$ because faint extended flux may be spreaded over
many beams.

\subsection{Discussion of Individual Sources}

S5~0454+844:  This object is unresolved in 8.46 GHz and 14.94 GHz maps to the 0.1 and 0.4 mJy beam\mo\ level
respectively.  It is also unresolved in a 1.4 GHz, A-array map to the 0.1 mJy beam\mo\ \citep{rec01}.  The
unresolved nature of this source is not unexpected due to a redshift lower limit of $z=1.340$ that is based upon
a MgII absorption system \citep{sto97}.  

PKS~0735+178:  This object has a intervening
absorption system at $z=0.424$ \citep{car74,rec01}.  The 8.46 GHz and spectral index maps are overlaid in
Figure~\ref{fig-1}.  The 14.94 GHz map is shown in Figure~\ref{fig-2}.  The 8.46 GHz map shows an unusual jet-like
morphology to the south and west.  The spectral index map confirms that PKS~0735+178 consists of a flat-spectrum
core (\alphaxu\ $= -0.24$) and steep-spectrum (\alphaxu\ $\la -1$) jets, not multiple images of the core.  

\begin{figure}
\plotone{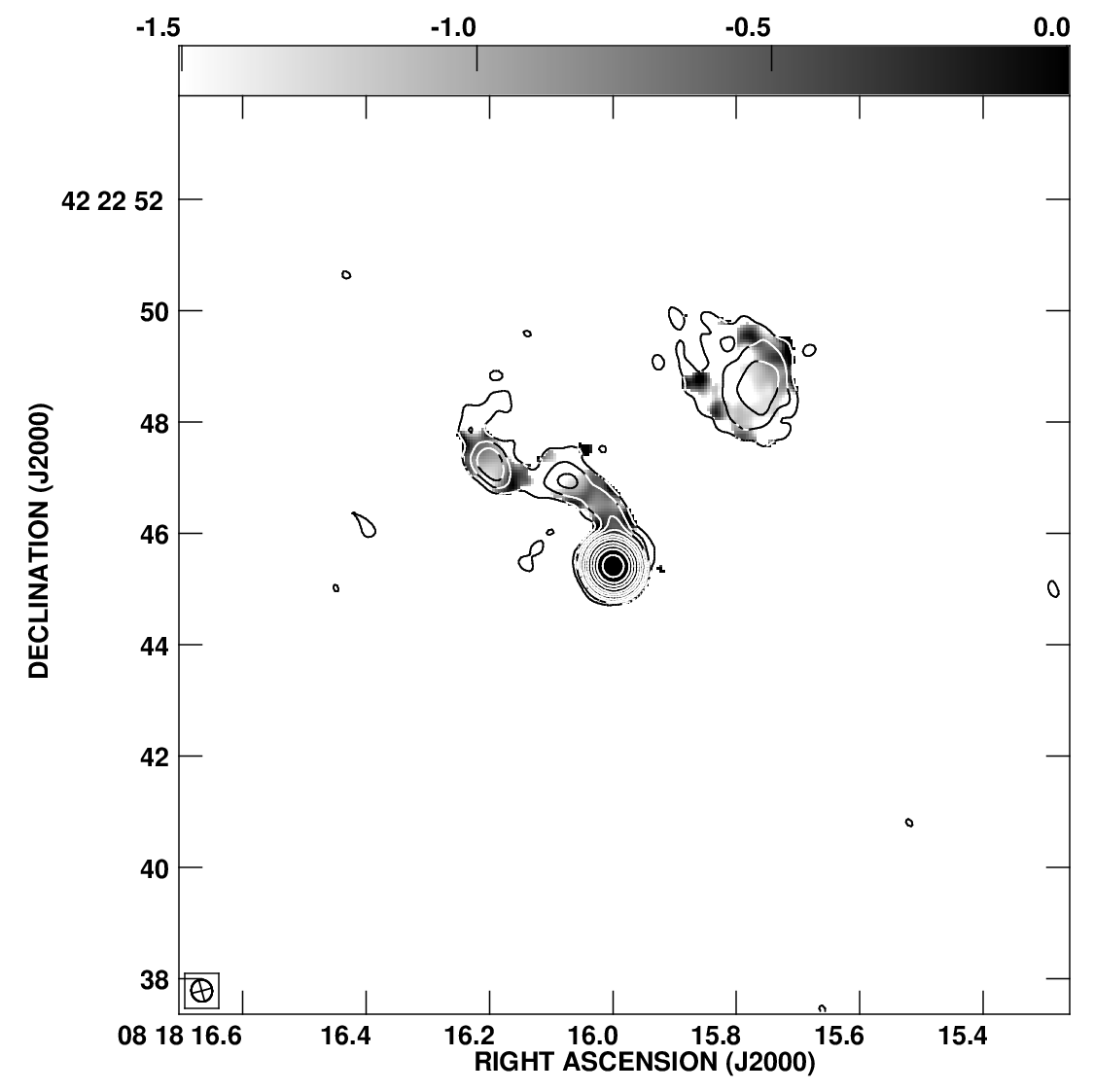}
\caption {VLA 4.86 GHz map (contour) and spectral index map (grayscale) of S4~0814+425.  The beam is shown in
the lower left corner.  The contour levels are 0.02, 0.05, 0.1, 0.2, 0.5, 1, 2, 5, 10, 20, 50 and 100\% the
peak flux of 1.0119 Jy beam\mo.  The spectral index range is from $-1.5$ to $0$.}
\label{fig-3}
\end{figure}

S4~0814+425:  The 1.4 GHz map of this object in \citet{mur93} is highly suggestive of a lensed system,
with possible multiple images of the core to the northeast and northwest.  However, 4.86 GHz and 8.46 GHz maps of
this object (Figure~\ref{fig-3} and Figure~\ref{fig-4}) show a jet which originates north of the core and
arcs to the northeastern lobe.  The spectral index map (Figure~\ref{fig-3}) confirms that the source consists
of a flat-spectrum core (\alphacx\ $= +0.1$) and steep-spectrum (\alphacx\ $\la -1$) jets.  The morphology of
this source is very unusual; however projection effects may be causing an intrinsically straight jet appear to
bend dramatically.

PKS~0823+033:  This object is unresolved in both the 8.46 GHz and 14.94 GHz maps to the 0.3 and 0.8 mJy  
beam\mo\ level respectively.  The object is also unresolved in the 1.4 GHz, A-array map in
\citet{mur93}.

PKS~1749+096:  This object is unresolved in both the 8.46 GHz and 14.94 GHz maps to the 0.8 and 0.9 mJy  
beam\mo\ level respectively.  It is also unresolved in a 1.4 GHz, A-array map to the 0.1 mJy beam\mo\
\citep{rec01}.  The modest tentative redshift ($z=0.320$:) and the fact that the host galaxy is easily observable
\citep{wur97} make its very high radio core dominance very unusual.  However, the observational limits allow that
this source could be a typical FR--1 with a somewhat lower extended radio power level (log
$P_{\rm 20 cm} \la 24.5$ W Hz\mo).

\begin{figure}
\plotone{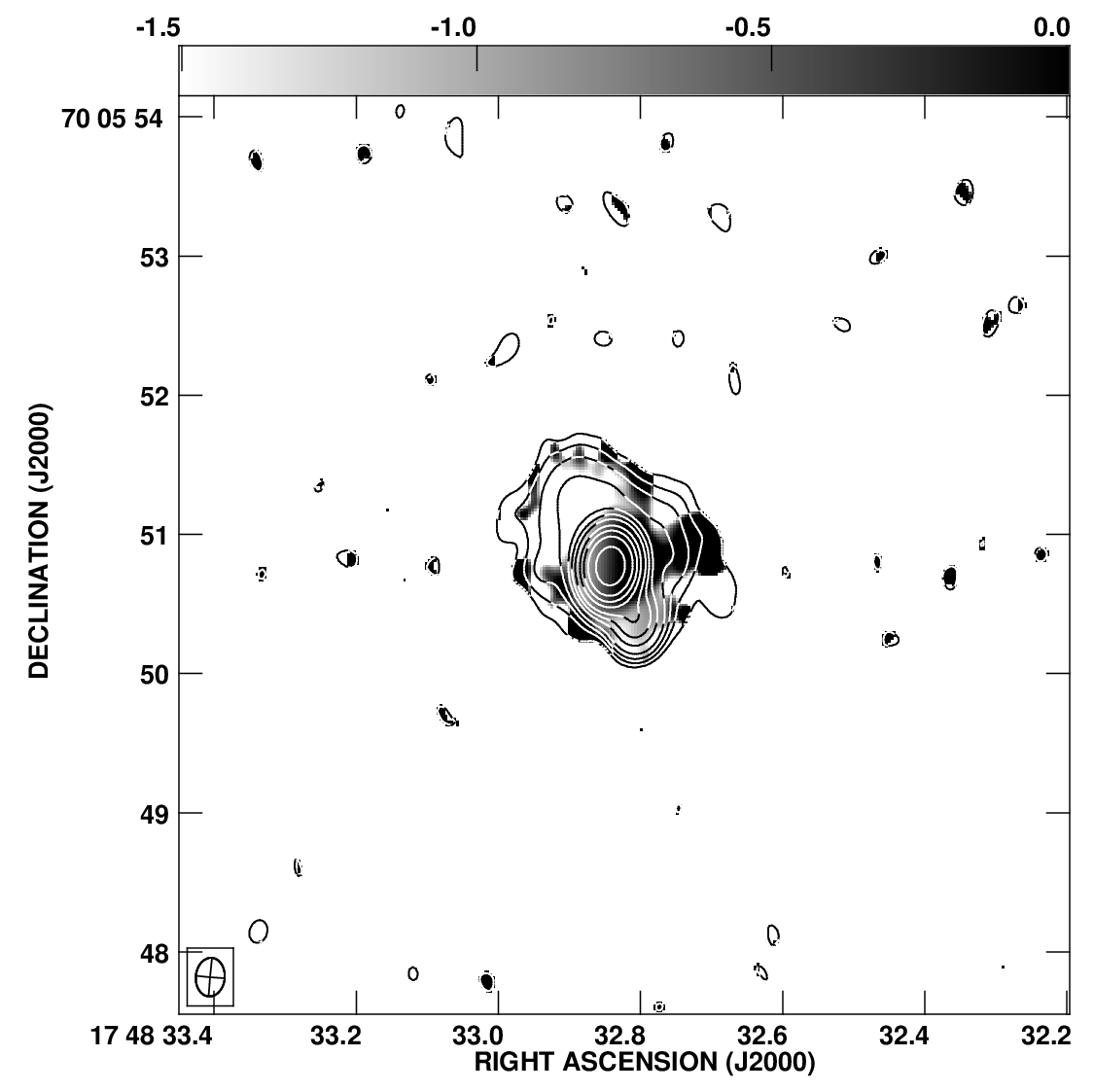}
\caption {VLA 8.46 GHz map (contour) and spectral index map (grayscale) of S4~1749+701.  The beam is shown in
the lower left corner.  The contour levels are 0.02, 0.05, 0.1, 0.2, 0.5, 1, 2, 5, 10, 20, 50 and 100\% the
peak flux of 5.209 x 10$^{-1}$ Jy beam\mo.  The spectral index range is from $-1.5$ to $0$.}
\label{fig-5}
\end{figure}

S4~1749+701:  This object was slightly resolved in 1.4 GHz map \citep{rec01}.  The 8.46 GHz map
(Figure~\ref{fig-5}) shows a halo morphology around the core with a jet-like extension to the southwest.
The halo is mostly resolved out of the 14.94 GHz map (Figure~\ref{fig-6}) however the extension to the
southwest is still visible.  The spectral index map (Figure~\ref{fig-5}) shows a a flat-spectrum core 
(\alphaxu\ $= -0.2$) and steep-spectrum (\alphaxu\ $\la -1$) halo, indicating that this halo is not an Einstein
ring image of the core.

\begin{figure}
\plotone{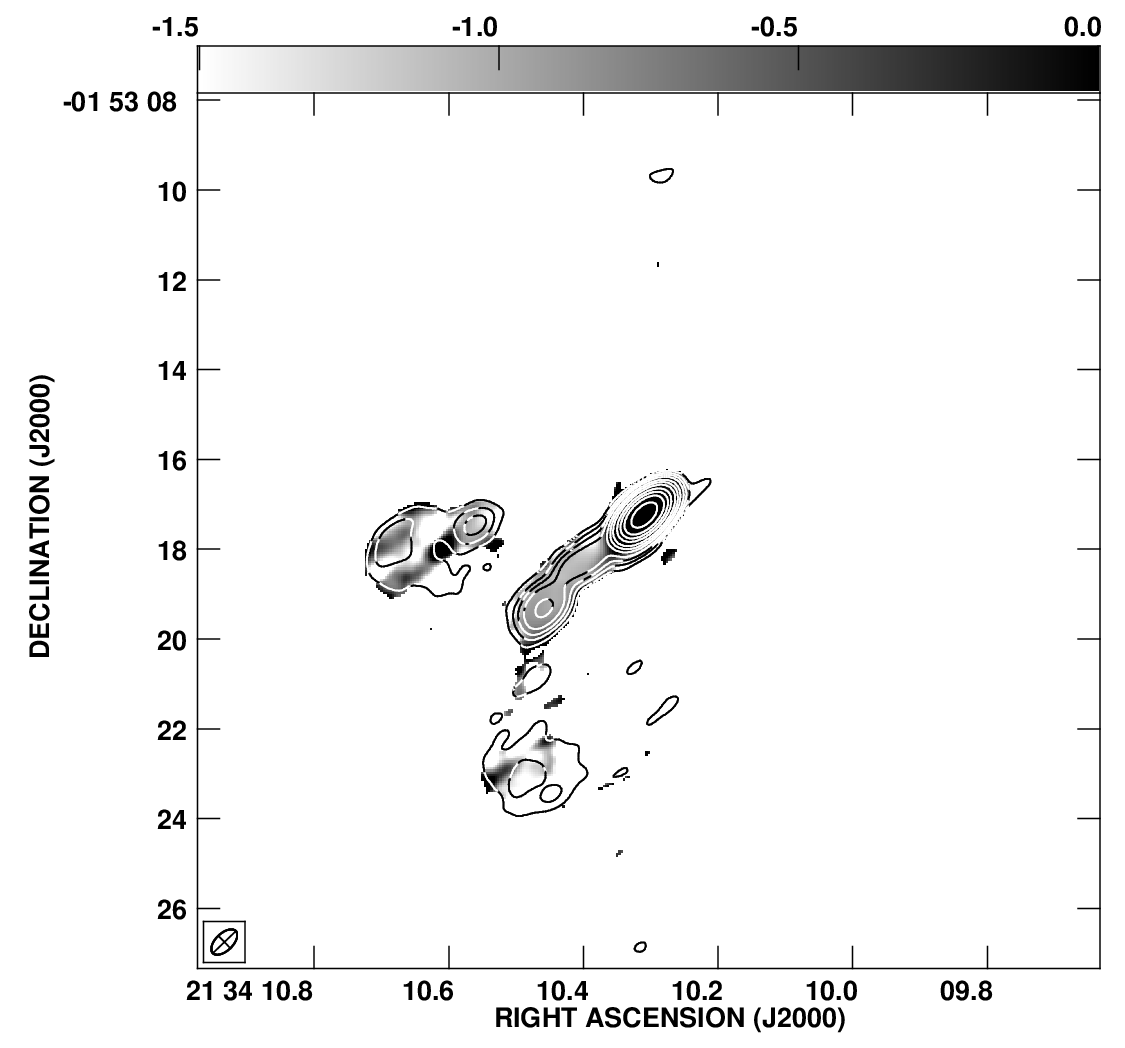}
\caption {VLA 4.86 GHz map (contour) and spectral index map (grayscale) of PKS~2131--021.  The beam is shown in
the lower left corner.  The contour levels are 0.02, 0.05, 0.1, 0.2, 0.5, 1, 2, 5, 10, 20, 50 and 100\% the
peak flux of 1.7321 Jy beam\mo.  The spectral index range is from $-1.5$ to $0$.}
\label{fig-7}
\end{figure}

PKS~2131--021:  The 1.4 GHz map of this object in \citet{rec01} is highly suggestive of a lensed system,
with possible multiple images of the core to the east and southeast.  However, 4.86 GHz and 8.46 GHz maps of
this object (Figures~\ref{fig-7} and~\ref{fig-8}) show a very interesting morphology, with a jet that
originates to the south-southeast of the core.  The jet appears to be significantly distorted and may be helical.  There also appears to be lobes roughly 5\arcsec\ to the east and southeast of the core.  It is not clear
how these jets propagate to the lobes, although projection effects may cause a single lobe to appear as the two
seen.  But significant bending is still required, even with the severe projection effects expected in BL Lacs,
to cause both lobes to appear on the same side of the core.  The spectral index map (Figure~\ref{fig-7}) shows no
evidence of multiple images of the core; and the jet and lobes have indices (\alphacx\ $\la -1$) much steeper than
the core (\alphacx\ $= +0.06$).
A MERLIN+EVN map of this object shows a jet component 50 mas east-southeast from the core, and multiple components extending to the south out to 350 mas from the core, suggesting a helical motion \citep{cas02}. 

\section{Discussion and Conclusions}

We have presented high-resolution radio continuum and spectral index maps for seven gravitational lensing
candidates from the 1Jy BL Lac sample.  All of the resolved sources show a flat-spectrum core and steep-spectrum extended radio morphology, as expected for core-jet sources.  We note that S4~0814+425 and PKS~2131--021 have very similar, and very unusual, radio morphologies in that both appear to have radio lobes on the same side of the core.  This could be caused by an intrinsically bent source seen close to the line of sight, although the required geometry would be rare.  If so, these sources should have very high apparent luminosities.  This is certainly the case for  PKS~2131--021, which has the highest core and extended radio powers of any BL Lac object in the 1Jy sample, discounting uncertain values \citep{rec01}.  The redshift for S4~0814+425 is currently unknown, although it must be $z\geq 0.5$ because it is unresolved at optical wavelengths \citep{pur02,urr99}; thus it too must have a high apparent radio luminosity.  VLBI maps of these sources indicate they are highly core-dominated, supporting the hypothesis they are highly beamed \citep{gab00,cas02}.  Unfortunately, further comparison is impossible without a firm redshift for S4~0814+425.

None of the maps show evidence of an Einstein ring or
multiply-imaged components; however it is important to note that a foreground galaxy many not cause
macrolensing if it has a low surface mass density \citep{nar90}.  Thus, it is possible that a foreground galaxy
may cause significant microlensing, brightening the continuum relative to the emission-line regions, without
associated macrolensing.  The converse is also true
\citep{abr93}.  Thus, these observations do not rule out the possibility of lensing; but they argue strongly
against macrolensing.  The resolved sources show highly distorted jets and misaligned lobes.  These unusual
morphologies support the hypothesis that these BL Lacs are highly beamed sources, with radio morphologies that are
highly distorted due to projection effects.

\acknowledgments

\clearpage


%
%




\clearpage

\begin{deluxetable}{rrrrr}
\tablecaption{Gravitational Lens Candidates \label{tbl-0}}
\tablewidth{0pt}
\tablehead{
\colhead{Object} &  \colhead{$z_{em}$} &  \colhead{$z_{abs}$} &  \colhead{log $P_{ext}$} & 
\colhead{$f$} }
\startdata
S5~0454+844   & \nodata & 1.340   & $\geq 27.31$ & $> 55.5$  \\
PKS~0735+178  & \nodata & 0.424   & $\geq 27.30$ & 231       \\
S4~0814+425   & \nodata & \nodata & \nodata      & $> 116.5$ \\
PKS~0823+033  & 0.506:  & \nodata & 27.31 & 151.7   \\
PKS~1749+096  & 0.320:  & \nodata & 26.59 & $> 145$   \\
S4~1749+701   & 0.770:  & \nodata & 27.23 & 15.8   \\
PKS~2131--021 & 1.285   & \nodata & 27.99 & 6.9   \\

\enddata
\end{deluxetable}

\clearpage

\begin{deluxetable}{rrrrrrr}
\tablecaption{Observed Radio Properties of the Sample \label{tbl-1}}
\tablewidth{0pt}
\tablehead{
\colhead{} & \multicolumn{2}{c}{4.86 GHz} & \multicolumn{2}{c}{8.46 GHz} & \multicolumn{2}{c}{14.94 GHz} \\
\cline{2-3} \cline{4-5} \cline{6-7} \\
\colhead{Object} & \colhead{$f_{core}$ (mJy)} & \colhead{$f_{ext}$ (mJy)} &
\colhead{$f_{core}$ (mJy)} & \colhead{$f_{ext}$ (mJy)} &
\colhead{$f_{core}$ (mJy)} & \colhead{$f_{ext}$ (mJy)} 
}
\startdata
S5~0454+844   & \nodata & \nodata & 233.3   & $< 0.5$      & 216.5   & $< 1.5$  \\
PKS~0735+178  & \nodata & \nodata & 975.5   & $32.2\pm3.3$ & 838.9   & $29.3\pm6.6$  \\
S4~0814+425   & 1011.9  & $33.0\pm4.2$     & 1072.3  & $18.4\pm6.7$          & \nodata & \nodata  \\
PKS~0823+033  & \nodata & \nodata & 1452.9  & $< 1.4$      & 1547.6  & $< 1.8$  \\
PKS~1749+096  & \nodata & \nodata & 3794.6  & $< 2.0$      & 3622.3  & $< 2.7$  \\
S4~1749+701   & \nodata & \nodata & 520.9   & $31.0\pm1.1$          & 459.9   & $18.6\pm5.8$      \\
PKS~2131--021 & 1732.1  & $70.8\pm8.0$     & 1802.0  & $36.3\pm6.7$          & \nodata & \nodata  \\

\enddata
\end{deluxetable}

\clearpage

\clearpage

\begin{thebibliography}{}

\bibitem[Abraham et al.(1993)]{abr93} Abraham, R.G., Crawford, C.S., Merrifield, M.R., Hutchings, J.B. \&
McHardy, I.M. 1993 \apj\ 415, 101.
\bibitem[Antonucci \& Ulvestad (1985)]{ant85} Antonucci, R.R.J. \& Ulvestad, J.S. 1985 \apj\ 294, 158.
\bibitem[Blandford \& Rees(1978)]{bla78} Blandford, R. \& Rees, M.J. 1978 in Proc. Pittsburgh Conference on
BL Lac Objects, ed. A.N. Wolfe, p. 328.
\bibitem[Briggs(1995)]{bri95} Briggs, D. 1995, PhD dissertation, New Mexico Institute of Mining and
Technology.
\bibitem[Carswell et al.(1974)]{car74} Carswell, R.F., Strittmatter, P.A., Williams, R.E., Kinman, T.D. \&
Serkowski, K. 1974 \apj\ 190, L101.
\bibitem[Cassaro et al.(1999)]{cas99} Cassaro, P., Stanghellini, C., Bondi, M., Dallacasa, D., della Ceca, R. \&
Zappal\`a, R.A. 1999 \aaps\ 139, 601.
\bibitem[Cassaro et al.(2002)]{cas02} Cassaro, P., Stanghellini, C., Dallacasa, D., Bondi, M. \& Zappal\`a, R.A. 2002 \aap\ 381, 378.
\bibitem[Dyer \& Shaver(1992)]{dye92} Dyer, C.C. \& Shaver, E.G. 1992 \apj\ 390, L5.
\bibitem[Falomo, Melnick \& Tanzi(1992)]{fal92} Falomo, R., Melnick, J. \& Tanzi, E.G. 1992 \aap\ 255L, 17.
\bibitem[Falomo et al.(1997)]{fal97} Falomo, R., Kotilainen, J., Pursimo, T., Sillanp\"a\"a, A., Takalo, L.
\& Heidt, J. 1997 \aap\ 321, 374. 
\bibitem[Fanaroff \& Riley(1974)]{fan74} Fanaroff, B.L. \& Riley, J.M. 1974 \mnras\ 167, 31.
\bibitem[Gabuzda, Pushkarev \& Cawthorne(2000)]{gab00} Gabuzda, D.C., Pushkarev, A.B. \& Cawthorne, T.V. 2000 \mnras\ 319, 1109.
\bibitem[Heidt et al.(1998)]{hei98} Heidt, J. et al. 1998, in Proc. Turku Conference on BL Lac Phenomenon,
ed. L. Takalo.
\bibitem[King et al.(1997)]{kin97} King, L.J. et al. 1997 \mnras\ 289, 450.
\bibitem[Murphy, Browne \& Perley(1993)]{mur93} Murphy, D.W., Browne, I.W.A. \& Perley, R.A. 1993 \mnras\
264, 298.
\bibitem[Nilsson et al.(1999)]{nil99} Nilsson, K., Takalo, L.O., Pursimo, T., Sillanp\"a\"a, A.,
Heidt, J., Wagner, S.J., Laurent-Muehleisen, S.A. \& Brinkmann, W. \aap\ 343, 81.
\bibitem[Narayan \& Schneider(1990)]{nar90} Narayan, R. \& Schneider, P. 1990 \mnras\ 243, 192.
\bibitem[O'Dea et al.(1992)]{ode92} O'Dea, C.P. et al. 1992 \aj\ 104, 1320.
\bibitem[Ostriker \& Vietri(1990)]{ost90} Ostriker, J.P. \& Vietri, M. 1990 Nature 344, 45.
\bibitem[Padovani \& Urry(1990)]{pad90} Padovani, P. \& Urry, C.M. 1990 \apj\ 356, 75.
\bibitem[Padovani \& Giommi(1995)]{pad95} Padovani, P. \& Giommi, P. 1995 \apj\ 444, 567.
\bibitem[Perlman \& Stocke(1994)]{per94} Perlman, E.S. \& Stocke, J.T. 1994 \aj\ 108, 56.
\bibitem[Perlman et al.(1996)]{per96} Perlman, E.S., Carilli, C.L., Stocke, J.T. \& Conway, J. 1996 \aj\
111,1839.
\bibitem[Pursimo et al.(2002)]{pur02} Pursimo, T., Nilsson, K., Takalo, L.O., Sillanp\"a\"a, A., Heidt, J. \&
Pietilae, H. 2002 \aap\ 381, 810.
\bibitem[Rector et al.(2000)]{rec00} Rector, T.A., Stocke, J.T., Perlman, E.S., Morris, S.L. \& Gioia, I.M. 2000
AJ 120, 1626.
\bibitem[Rector \& Stocke(2001)]{rec01} Rector, T.A. \& Stocke, J.T. 2001 \aj\ 122, 565.
\bibitem[Romero, Surpi \& Vucetich(1995)]{rom95} Romero, G.E., Surpi, G. \& Vucetich, H. 1995 \aap\ 301, 64.
\bibitem[Scarpa et al.(1999)]{sca99} Scarpa, R., Urry, C.M., Falomo, R., Pesce, J.E., Webster, R., O'Dowd, M.
\& Treves, A. 1999 \apj\ 521, 134.
\bibitem[Scarpa et al.(2000)]{sca00} Scarpa, R., Urry, C.M., Falomo, R., Pesce, J.E. \& Treves, A.
2000 \apj\ 532, 740. 
\bibitem[Steidel \& Sargent(1992)]{ste92} Steidel, C.C. \& Sargent, W.L.W. 1992 \apjs\ 80, 90.
\bibitem[Stocke, Wurtz \& Perlman(1995)]{sto95} Stocke, J.T., Wurtz, R.E. \& Perlman, E.S. 1995 \apj\ 454, 55.
\bibitem[Stocke \& Rector(1997)]{sto97} Stocke, J.T. \& Rector, T.A. 1997 \apj\ 489, L17.
\bibitem[Stickel, Fried \& K\"uhr(1988)]{sti88} Stickel, M., Fried, J.W. \& K\"uhr, H. 1988 \aap\ 198, L13.
\bibitem[Stickel et al.(1991)]{sti91} Stickel, M., Padovani, P., Urry, C.M., Fried, J.W. \& K\"uhr,
H. 1991 \apj\ 374, 431.
\bibitem[Takalo et al.(1998)]{tak98} Takalo, L. et al. 1998 \aaps\ 129, 577.
\bibitem[Urry \& Padovani(1995)]{urr95} Urry, C.M. \& Padovani, P. 1995 \pasp\ 107, 803.
\bibitem[Urry et. al(1999)]{urr99} Urry, C.M., Falomo, R., Scarpa, R., Pesce, J.E., Treves, A. \& Giavalisco, M. 1999 \apj\ 512, 88.
\bibitem[Webb et al.(2000)]{web00} Webb, J.R., Howard, E., Ben\'itez, E., Balonek, T., McGrath, E., Shrader, C.,
Robson, I. \& Jenkins, P. 2000 \aj\ 120, 41.
\bibitem[Wurtz et al.(1997)]{wur97} Wurtz, R., Stocke, J.T. \& Yee, H.K.C. 1997 \apj\ 480, 5.


\end{thebibliography}
\end{document}